\documentclass[english,12pt]{article}
\usepackage{array}
\usepackage{graphicx}
\usepackage{amssymb}
\usepackage{amsmath}
\usepackage{multirow}
\usepackage{prettyref}
\usepackage{babel}
\usepackage{units}
\usepackage[latin1]{inputenc}
\usepackage{amsfonts}
\usepackage{amssymb}
\usepackage{babel}
\usepackage{subcaption}
\usepackage{cite}
\usepackage{xcolor}
\def\@fmsl@sh#1#2#3{\m@th\ooalign{$\hfil#1\mkern#2/\hfil$\crcr$#1#3$}}
 \def\eq#1\en{\begin{equation}#1\end{equation}}
\def\s[#1,#2]{[#1\stackrel{\star}{,}#2]}
\def\sx[#1,#2]{[#1\stackrel{\star_{x}}{,}#2]}

\newcommand{\nc}{\newcommand}
\nc{\beq}{\begin{equation}}
\nc{\eeq}{\end{equation}}
\nc{\beqa}{\begin{eqnarray}}
\nc{\eeqa}{\end{eqnarray}}

\def\bc{\begin{center}}
\def\ec{\end{center}}

\def\to{\rightarrow}

\def\gsim{\mathrel{\mathpalette\atversim>}}

\def\bc{\begin{center}}
\def\ec{\end{center}}

\def\gsim{\mathrel{\rlap{\lower4pt\hbox{\hskip1pt$\sim$}}

    \raise1pt\hbox{$>$}}}       

\def\gsim{\mathrel{\rlap{\lower4pt\hbox{\hskip1pt$\sim$}}
    \raise1pt\hbox{$>$}}}       

\begin{document}
\makeatletter
\def\fmslash{\@ifnextchar[{\fmsl@sh}{\fmsl@sh[0mu]}}
\def\fmsl@sh[#1]#2{%
  \mathchoice
    {\@fmsl@sh\displaystyle{#1}{#2}}%
    {\@fmsl@sh\textstyle{#1}{#2}}%
    {\@fmsl@sh\scriptstyle{#1}{#2}}%
    {\@fmsl@sh\scriptscriptstyle{#1}{#2}}}
\def\@fmsl@sh#1#2#3{\m@th\ooalign{$\hfil#1\mkern#2/\hfil$\crcr$#1#3$}}
\makeatother

\thispagestyle{empty}
\begin{titlepage}
\boldmath
\begin{center}
  \Large {\bf Exorcising ghosts in quantum gravity}
    \end{center}
\unboldmath
\vspace{0.2cm}
\begin{center}
{{\large Iber\^e Kuntz$^{abc}$}\footnote{kuntz@bo.infn.it}}
 \end{center}
\begin{center}
{\sl $^a$Federal University of ABC, Center of Mathematics,  Santo Andr\'e, Brazil}
\\
$ $
\\
{\sl
$^b$
Dipartimento di Fisica e Astronomia, Universit\`a di Bologna,
\\
via Irnerio~46, I-40126 Bologna, Italy}
\\
$ $
\\
{\sl 
$^c$
I.N.F.N., Sezione di Bologna, IS - FLAG
\\
via B.~Pichat~6/2, I-40127 Bologna, Italy
}
\end{center}
\vspace{5cm}
\begin{abstract}
\noindent We remark that Ostrogradsky ghosts in higher-derivative gravity, with a finite number of derivatives, are fictitious as they result from an unjustified truncation performed in a complete theory containing infinitely many curvature invariants. The apparent ghosts can then be projected out of the quadratic gravity spectrum by redefining the boundary conditions of the theory in terms of an integration contour that does not enclose the ghost poles. This procedure does not alter the renormalizability of the theory. One can thus use quadratic gravity as a quantum field theory of gravity that is both renormalizable and unitary.
\end{abstract}  
\end{titlepage}



\newpage

Despite the major advances in the quantization of gravity obtained in the past few decades, a deep understanding of quantum gravity in the UV remains a matter of debate. General relativity is known to be non-renormalizable, generating higher curvature invariants in the action, which are required for renormalization \cite{tHooft:1974toh}. However, by introducing higher-derivative terms, ghosts inevitably appear in the spectrum unless the theory is treated under the effective field theory formalism where the higher derivatives are seen as perturbations \cite{Simon:1990jn,Donoghue:1994dn}. The purpose of this paper is to remark that Ostrogradskian ghosts in higher-derivative gravity are only apparent when one truncates the infinite series of curvature invariants. We then show how these ghosts can be removed by means of a suitable boundary condition.

The known issue with higher powers of the curvature invariants is due to Ostrogradsky theorem \cite{Ostrogradsky:1850fid,Eliezer:1989cr,Woodard:2006nt}. It states that any dynamical system described by differential equations containing time derivatives higher than second (but finite) order necessarily possesses unbounded energy solutions, dubbed ghosts. The existence of a ghost is not itself an issue, but it becomes a problem when the ghost field interacts with other sectors, which allows for the endless process of transmitting energy from healthy fields to the ghost. At the quantum level, negative energy states are sometimes traded by states with negative norm which is again problematic as it violates the optical theorem \cite{Cline:2003gs}. One simple way of evading Ostrogradsky theorem is with degenerate theories, such as $f(R)$, but as we will see, functions of the Ricci scalar are not sufficient for renormalization \cite{Stelle:1976gc}. We will show yet another way of evading Ostrogradsky theorem in quantum gravity.

The idea of embedding higher-derivative gravity into a theory with infinite curvature invariants is not new. This idea has been in fact a steppingstone for the infinite derivative gravity, whose gravitational propagator is defined to be an entire function with a single pole at vanishing momentum \cite{Biswas:2005qr,Modesto:2011kw,Biswas:2011ar,Modesto:2017sdr}. This guarantees from the onset that no particle (ghost or otherwise) other than the graviton is present. Asymptotically safe gravity also benefit from the same idea of embedding a finite truncation of the action into a complete theory with infinitely many terms. It thus share the problems with ghosts, which could also be solved by an entire function extension of the propagator. In the asymptotic safety scenario, there is also the possibility that the ghost mass goes to infinity as the theory approaches the non-trivial fixed point \cite{Becker:2017tcx}. In this limit, the ghost would then decouple from the theory.

In this paper, we remark that any action consisting of infinitely many terms is free from Ostrogradskian ghosts. We must emphasize that this result abolishes only ghosts that result from the Ostrogradsky theorem, which justifies the name Ostrogradskian ghosts. There could be, nonetheless, other types of ghosts in the infinite-derivative theory, for which we shall reserve the adjective ``non-Ostrogradskian'', that must be dealt with in a different fashion, e.g. by forcing their masses to go to infinity or implementing entire functions in the propagator. In fact, it has been shown that some infinite-derivative theories are free of ghosts at the tree level, but ghosts can reappear when loop corrections are included \cite{Shapiro:2015uxa}. These types of ghosts obviously do not result from the Ostrogradsky theorem, otherwise they would be present from the onset in the classical action. Futhermore, as we shall see, the Ostrogradsky theorem does not hold for infinite-derivative theories. This does not imply, of course, the existence of a no-go theorem for non-Ostrogradskian ghosts. The failure of Ostrogradsky's theorem for infinite-derivative theories merely means that \textit{Ostrogradskian ghosts} (as opposed to other potential types of ghosts) are absent \footnote{By inspection, one can in fact always manufacture infinite-derivative actions with non-Ostrogadskian ghosts by, for example, multiplying the bilinear term in the action by an exponential form factor, but none of these ghosts originate from the Ostrogradsky theorem. Therefore, the statement that Ostrogradskian ghosts are absent for any theory containing infinitely many derivatives does not clash with the examples in the literature where non-Ostrogradskian ghosts are present \cite{Shapiro:2015uxa,Barnaby:2010kx}. Such examples, in fact, exploit avenues beyond the Ostrogradsky theorem, such as the creation of non-Ostrogradskian ghosts via radiative corrections.}. For this reason, we will adopt the nomenclature Ostrogradskian ghosts for those ghosts deduced from Ostrogradsky theorem and non-Ostrogradskian ghosts for any other type of ghost.

Classical general relativity is described by the Einstein-Hilbert action
\beq
S_{EH} = \int\mathrm{d}^4x \sqrt{-g} \, \frac{M_p^2}{2} R,
\eeq
where $M_p$ is the Planck mass. Upon quantization using e.g. the background field method, one finds that one-loop divergences are proportional to second order curvature invariants, i.e. terms containing four derivatives \cite{tHooft:1974toh}. In order to absorb these divergences, one must thus include counter-terms of the form:
\begin{equation}
\Delta \mathcal L_{1-loop} = a_1 R^2 + a_2 R_{\mu\nu}R^{\mu\nu} + a_3 R_{\mu\nu\rho\sigma}R^{\mu\nu\rho\sigma} + a_4 \Box R,
\label{eq:action}
\end{equation}
where $a_i$ are parameters chosen to cancel out the one-loop divergences. Similarly, the renormalization of general relativity at two-loop order requires terms of cubic order such as $R\Box R$ \cite{Goroff:1985th}, tree-loop corrections in turn requires fourth order counter-terms and so on. This process continues \textit{ad infinitum}, clearly indicating that quantum general relativity is not predictive. This is in fact a reflection of the non-renormalizability of general relativity as can be seen from the superficial degree of divergence $D$ of Feynman diagrams with $p$ loops and $d$ metric derivatives in the counter-terms \cite{Stelle:1976gc,Lavrov:2019nuz}:
\beq
D + d = 4 + 2p,
\eeq
which increases without limit with the number of loops $p$. This problem is circumvented within the realm of effective field theories \cite{Donoghue:1994dn} (see \cite{Burgess:2003jk} for a review). In the effective field theory description of quantum gravity, terms in the action are organized in powers of $E/M_p$, where $E$ is the typical energy of the problem. Dimensional analysis shows that higher-order curvatures correspond to higher powers of the $E/M_p$, thus at energies way below the Planck scale the higher powers of the curvature are utterly small and can be treated as tiny perturbations. Thus at any given precision, the infinite series can be truncated, producing only a finite number of free parameters. In this scenario, there is no new degree of freedom, ghost or otherwise, besides the standard graviton and the interaction is that of general relativity. The higher-order terms capture the underlying physics perturbatively and only contribute to the vertices of Feynman diagrams, not to the propagators. As a result, one obtains a theory that can be renormalized, albeit being non-renormalizable, at every loop order without introducing ghosts to the spectrum, but that only makes sense at energies below the Planck scale.

On the other hand, we could let the fourth derivative terms in \eqref{eq:action} take arbitrary values which would make them compete with the Einstein-Hilbert term at the Planck scale \cite{Stelle:1976gc}. In this case, the bare action reads
\beq
S_{HD} = \int\mathrm{d}^4x \sqrt{-g} \, \left[\frac{M_p^2}{2} R + b_1 R^2 + b_2 R_{\mu\nu}R^{\mu\nu} + b_3 R_{\mu\nu\rho\sigma}R^{\mu\nu\rho\sigma} + b_4 \Box R \right],
\label{eq:hd}
\eeq
where $b_i$ are bare coupling constants. The higher-derivative terms now are not mere counter-terms as in general relativity (see Eq.~\eqref{eq:action}). They are treated on the same footing as the Einstein-Hilbert term, thus contributing to both vertices and propagators. This theory, which obviously differs from quantum general relativity, came to be known as higher-derivative gravity. Albeit the modification with respect to quantum general relativity seems minimal, the change in the divergent structure is drastic. In fact, the superficial degree of divergence for Eq.~\eqref{eq:hd} reads \cite{Stelle:1976gc,Lavrov:2019nuz}
\beq
D + d = 4 - \sum_{r=0}^1(4-2r)n_{2r},
\label{eq:DHD}
\eeq
where $d$ is the number of metric derivatives in the counter-terms and $n_{2r}$ is the number of graviton vertices with $2r$ derivatives. Since \eqref{eq:DHD} does not depend on the number of loops $p$, the action with second order curvature invariants \eqref{eq:hd} is renormalizable to all loop orders and one need not include terms with even higher derivatives \footnote{Derivatives higher than fourth order can make the theory superrenormalizable \cite{Asorey:1996hz} and unitary at the tree level \cite{Modesto:2015ozb}.}. This theory could be interpreted as a fundamental theory for quantum gravity if it was not for the presence of a ghost in its spectrum \cite{Stelle:1976gc}. Unitarity can thus be traded by renormalizability. Higher-derivative gravity has the advantage of having interesting new solutions, such as Starobinsky inflation \cite{Starobinsky:1980te}, but its drawback is the instability and the violation of unitarity caused by the ghost\footnote{Note that Starobinsky inflation is based on the Lagrangean $R+R^2$, which is ghost-free but non-renormalizable. The additional term $Ric^2$ is necessary to obtain a renormalizable theory, but it is the culprit for the presence of the ghost. Note also that $Ric^2$ only contributes to the spin-2 sector, thus one can still have Starobinsky solution in some regimes as pointed out in \cite{Calmet:2019tur,Calmet:2016fsr}, which comes from the scalar field hidden in $R^2$.}. Some solutions to the non-unitarity problem have been proposed \cite{Lee:1969fy,Cutkosky:1969fq,Tomboulis:1977jk,Tomboulis:1983sw}, but a consistent way of dealing with the classical ghost instabilities remains unknown. We shall argue that this ghost is however absent when one takes into account all possible curvature invariants. We must emphasize that non-Ostrogradskian ghosts generally exist in a theory with infinite derivatives. Nonetheless, we are interested in justifying the removal of the Ostrogradsky ones, namely those resulting from Ostrogradsky theorem, since these are the ones that show up in quadratic gravity, which is ultimately the theory we are interested in.

We have seen that higher-derivative gravity must contain fourth derivative terms in the action for renormalizability. Nonetheless, nothing forbids us from including sixth derivative terms or higher, such as curvature terms of third order. In fact, there are infinitely many terms that are allowed by the diffeomorphism symmetry and there is a priori no reason to exclude them from the action. Furthermore, effective field theory requires the inclusion of $n$th-order curvature invariants to renormalize calculations with $n-1$ loops. Since effective field theory is a model-independent formulation of quantum gravity in the IR, any respectful quantum theory of gravity should give rise to (local and non-local) operators of arbitrary order in the low-energy regime, unless there is some additional symmetry that forbids the appearance of these terms. Therefore, even though renormalization only requires second order curvature invariants, higher-order operators are necessary for a consistent matching with the effective theory.

The subset of terms appearing in \eqref{eq:hd} is thus obtained by truncating the following infinite series of curvature invariants:
\begin{equation}
S_\infty = \lim_{N\to\infty} \sum_{n=0}^{N}\int\mathrm{d}^4x\sqrt{-g} \, \mathcal O_{2n},
\label{eq:full}
\end{equation}
where $\mathcal O_{2n}$ denotes a linear combination of all curvature invariants containing $2n$ derivatives. The coupling constants are included inside the definition of $\mathcal O_{2n}$. To second order, one thus finds:
\begin{align}
\mathcal{O}_2 &= \frac{M_p^2}{2} R,\\
\mathcal{O}_4 &= a_1 R^2 + a_2 R_{\mu\nu}R^{\mu\nu} + a_3 R_{\mu\nu\rho\sigma}R^{\mu\nu\rho\sigma} + a_4 \Box R.
\end{align}
In effective field theory, this truncation is well-defined and can be performed at any order without introducing ghosts or any other particle. This is possible because the set of operators $\mathcal O_{2n}$ becomes increasingly smaller as we crank up $n$. But in higher-derivative gravity, such a truncation is poorly motivated, although it is sometimes assumed that a theory of the type \eqref{eq:hd} could arise in one of the many vacua of string theory. We now show that the Ostrogradsky ghost is actually a result of this poor truncation.

A crucial step for proving Ostrogradsky theorem on instability consists in writing an $N$th-order differential equation
\begin{equation}
f(t,\dot x,\ddot x,\ldots,x^{(N)})=0
\end{equation}
as a system of $N$ coupled first-order differential equations
\begin{align}
\dot x_1(t) & = f_1(t, x_1, x_2, \ldots, x_N),\nonumber\\
\dot x_2(t) & = f_2(t, x_1, x_2, \ldots, x_N),\nonumber\\
&\vdots\nonumber\\
\dot x_N(t) & = f_N(t, x_1, x_2, \ldots, x_N),
\end{align}
where $x_n = x^{(n-1)}$ are phase space coordinates and $f_n$ can be expressed in terms of derivatives of the Hamiltonian $\mathcal H(t,x_1, x_2, \ldots, x_N)$. In this situation, the Hamiltonian $\mathcal H$ turns out to depend linearly on $N/2-1$ of its arguments, signaling that $\mathcal H$ is not positive-definite for a large portion of the phase space. However, as noted in the mathematical literature \cite{Lalesco,book1,Barnaby:2007ve}, an infinite differential equation cannot be written as a system of $N=\infty$ first-order differential equations by simply applying the same reasoning of the case where $N$ is finite. In particular, the Hamiltonian system for $N=\infty$ does not correspond to the same problem described by an infinite differential equation, thus one should not expect Ostrogradsky instabilities for differential equations of infinite order. Since the theory \eqref{eq:full} contains infinitely many higher-order operators in its action, its equation of motion is described by a differential equation of infinite order. We thus conclude that Ostrogradsky ghosts are necessarily absent in higher-derivative gravity when one includes every possible curvature invariant to the action. Evading Ostrogradsky theorem, however, does not guarantee that other types of ghosts will be absent. Nonetheless, it shows that the Ostrogradsky ghost present in quadratic gravity (Eq.~\eqref{eq:hd}) is an artificial particle that originates from the truncation. We should point out that there are theories whose ambition is to eliminate every ghost, Ostrogradskian or not, from the spectrum \cite{Tomboulis:1997gg,Modesto:2011kw}.

As a concrete example, consider quantum gravity in two dimensions as described by the Polyakov action \cite{Polyakov:1987zb}
\begin{equation}
S = \frac{1}{96\pi}\int\mathrm{d}^2x\sqrt{-g} R\frac{1}{\Box}R.
\label{eq:poly}
\end{equation}
The non-local operator $1/\Box$ can be seen as an infinite series of diffeomorphism invariants. In fact, one can bring \eqref{eq:poly} to the form
\begin{equation}
S = \frac{1}{96\pi}\int\mathrm{d}^2x\sqrt{-g}\sum_{n=0}^\infty R\frac{(-1)^n}{\Lambda^2} \left(\frac{\Box}{\Lambda^2} - 1\right)^n R,
\label{eq:polyinf}
\end{equation}
in the region of convergence of the series, where $\Lambda$ is an arbitrary expansion point with dimension of mass. The action \eqref{eq:poly} contains only a healthy scalar degree of freedom in its spectrum. This can be seen by performing a conformal transformation $g_{\mu\nu} = e^{2\sigma}\eta_{\mu\nu}$, which leads to \cite{Belgacem:2017cqo}
\begin{equation}
S = \int\mathrm{d}^2x \frac{-1}{24\pi}\eta^{\mu\nu}\partial_\mu\sigma\partial_\nu\sigma.
\end{equation}
Therefore, the infinite action \eqref{eq:polyinf} does not contain any ghost resulting from the Ostrogradsky construction. Any truncation of \eqref{eq:polyinf}, on the other hand, produces a higher-derivative theory of \textit{finite-order}, which is susceptible to the application of the Ostrogradsky theorem and thus contains Ostrogradskian ghosts in its spectrum. However, since these ghosts result from the artificial procedure of truncation, they are not genuine degrees of freedom. Another example is given by Barvinsky's non-local theory \cite{Barvinsky:2011rk}:
\begin{equation}
S = \frac{M^2}{2}\int\mathrm{d}^4x \sqrt{-g}\left[-R + \alpha R^{\mu\nu}\frac{1}{\Box + \hat P}G_{\mu\nu}\right],
\label{eq:barv}
\end{equation}
where $M$ is a dimensionful constant, $\alpha$ is a dimensionless parameter, $G_{\mu\nu}$ is the Einstein tensor and
\beq
\hat P \equiv P_{\alpha\beta}^{\ \ \mu\nu} = c_1 R_{(\alpha\ \ \beta)}^{\ \ (\mu\ \ \nu)} + c_2 (g_{\alpha\beta} R^{\mu\nu} + g^{\mu\nu} R_{\alpha\beta}) + c_3 R^{(\mu}_{(\alpha}\delta^{\nu)}_{\beta)} + c_4 R g_{\alpha\beta} g^{\mu\nu} + c_5 R \delta^{\mu\nu}_{\alpha\beta},
\eeq
$c_i$ being dimensionless parameters. The condition
\beq
\frac{c_1}{3} - c_3 - 4 c_5 = \frac{2}{3}
\label{eq:cond}
\eeq
guarantees that the theory is ghost-free. The degrees of freedom of \eqref{eq:barv} in this case are the same ones of general relativity, thus there is no Ostrogradsky ghost in the spectrum, which corroborates our result. In fact, the non-local piece can again be viewed as an infinite series by expanding it about the point $\Box + \hat P = \Lambda^2$ to give
\beq
S = \frac{M^2}{2}\int\mathrm{d}^4x \sqrt{-g}\left[-R + \sum_n \alpha R^{\mu\nu} \frac{(-1)^n}{\Lambda^2} \left(\frac{\Box + \hat P}{\Lambda^2} - 1\right)^n G_{\mu\nu}\right],
\label{eq:barv}
\eeq
where $\Lambda$ has dimension of mass as before. Once again, although Eq.~\eqref{eq:barv} contains no ghosts, any truncation of it would produce ghosts by the Ostrogradsky theorem. This example also illustrates the possibility of having non-Ostrogradskian ghosts, as opposed to Ostrogradskian ones, should the condition \eqref{eq:cond} not be satisfied.

Therefore, the apparent Ostrogradsky ghost that appears in \eqref{eq:hd} is just a byproduct of the unjustified truncation of $S_\infty$. Note that truncating $S_\infty$ to a finite order $N$ yields many non-physical particles that are not present in the full theory because the truncation inevitably changes the pole structure of the propagator \cite{Barnaby:2007ve,Becker:2017tcx}. Along with these particles, many ghosts are expected to appear as a result of Ostrogradsky theorem. We emphasize that all these particles, regardless of being ghosts, are fictitious as they only appear as an artifact of the truncation. Thus whenever we want to truncate $S_\infty$, we can very well project out these unphysical poles by suitably choosing an integration contour to define the truncated theory. This is done by first linearizing the equations of motion of the truncated theory around Minkowski, which ultimately leads to the generalized wave equation for the transverse and traceless perturbation $h_{\mu\nu}$\footnote{We focus on the homogeneous equation for the sake of the argument, but the method can be easily generalized to the case where matter is present. See \cite{Barnaby:2007ve} for a detailed discussion of the initial value problem of infinite and finite higher-derivative differential equations.}:
\begin{equation}
\mathcal F(\Box)h_{\mu\nu} = 0,
\label{eq:eom}
\end{equation}
where $\mathcal F(\Box)$ is a generalized wave operator. We can write the perturbation in Fourier space as
\begin{equation}
h_{\mu\nu} = \oint_\mathcal{C} \mathrm{d}^4q e^{-iqx} \tilde h_{\mu\nu}(q),
\end{equation}
where $\mathcal C$ is a contour that is chosen according to the desired boundary conditions. The action of the pseudo-differential operator $\mathcal F(\Box)$ is defined in Fourier space, where Eq. \eqref{eq:eom} becomes
\begin{equation}
\oint_\mathcal{C} \mathrm{d}^4q e^{-iqx} \mathcal F(-q^2)\tilde h_{\mu\nu}(q)=0.
\end{equation}
Note that the boundary conditions, as well as the contour $\mathcal C$, are part of the definition of the theory. The most standard choice is a contour that encloses all zeros of \eqref{eq:eom} (or equivalently, all poles of the propagator) and yet satisfies Feynman boundary conditions. But that is not the only choice. We can, for example, choose $\mathcal C$ without enclosing the ghost poles, keeping them from appearing in the truncated theory \cite{Barnaby:2007ve,Kuntz:2019gup}. The zeros of $\mathcal F(-q^2)$ can be isolated by means of the Weierstrass factorization theorem
\begin{equation}
\mathcal F(-q^2) = q^2 e^{g(q^2)}\prod_{n=1}^{\infty}\left(1-\frac{q^2}{q_n^2}\right)\exp\left\{\left(\frac{q}{q_n}\right)^2+\frac{1}{2}\left(\frac{q}{q_n}\right)^4+\cdots+\frac{1}{\lambda_n}\left(\frac{q}{q_n}\right)^{2\lambda_n}\right\},
\label{eq:weier}
\end{equation}
where $g(q^2)$ is a holomorphic function, $\{\lambda_n\}$ is a sequence of integers and $q_n$ are the zeros of $\mathcal F(-q^2)$. From Cauchy theorem, we obtain the solution to \eqref{eq:eom} restricted to a contour $\mathcal C$ that does not enclose any ghost
\begin{equation}
h_{\mu\nu} = \sum_{n}\sum_{s=+,-} a^{n,s}_{\mu\nu} e^{-iq_n^sx},
\end{equation}
where $a^{n,s}_{\mu\nu}$ are polarization tensors and $(q_n^s)^2$ are the poles of $\mathcal F(-q^2)^{-1}$ located inside the contour $\mathcal C$, i.e. plane waves corresponding to ghost fields are absent. Given that Ostrogradsky ghosts are spurious as they only appear in the spectrum as an artifact of the truncation, it is natural to make this choice. Projecting out a ghost particle prohibits its corresponding field of having plane wave solutions, thus not leading either to instabilities in the classical theory or to unitarity issues at the quantum level. In particular, the ghost field cannot be written in terms of creation and annihilation operators, thus it never appears in asymptotic states and does not constitute a physical particle. Its only effect is to intermediate a repulsive Yukawa interaction that originates from the particular solution to the inhomogeneous Klein-Gordon equation, i.e. from the Green's function. Restricting to the quadratic truncation, the Yukawa interaction for a point particle of mass $M$ reads \cite{Stelle:1977ry}
\begin{equation}
V = -\frac{M}{3\pi M_p^2}\frac{e^{-m_2 r}}{r},
\end{equation}
where $m_2 = \sqrt{\frac{M_p^2}{2 a_2}}$. The truncated theory is then free of pathologies and yet able to capture the essence of quantum gravity.

A few comments are in order. Firstly, it is important to note that projecting out the ghost particle does not recover renormalization issues. In fact, the functional form of the gravitational propagator in quadratic gravity continues to behave as $q^{-4}$, namely
\beq
\Delta(-q^2) \sim \frac{1}{q^2} - \frac{1}{q^2 - m^2_2},
\eeq
since $\Delta(-q^2)$ is the inverse of the operator $\mathcal F(-q^2)$ regardless the contour $\mathcal C$. The second term in $\Delta(-q^2)$ is present for any choice of $\mathcal C$, allowing to cancel the divergences from (massless) graviton loops in analogy to the Pauli-Villars regularization. In particular, the superficial degree of divergence $D$ of Feynman diagrams remains that of Eq.~\eqref{eq:DHD}, indicating that the theory is indeed renormalizable. The ghost pole is however absent because we have changed the boundary conditions, namely the contour $\mathcal C$, and the momentum $q$ never hits the ghost pole $q^2 = m^2_2$. This means that the ghost cannot be put on-shell, thus it no longer appears in external legs of Feynman diagrams (asymptotic states). It is important to recall that only on-shell states contribute to the optical theorem. Therefore, the ghost can exist off-shell, which makes the theory renormalizable, but it never appears in asymptotic states, thus preventing unitarity issues\footnote{A similar approach consists in the quantization of ghosts as fakeons, which are also particles that only live off-shell \cite{Anselmi:2017ygm,Anselmi:2018kgz}.}. Note that although we have taken the quadratic truncation as our primary example, the above considerations extends to truncations of any finite order $N$. Secondly, the procedure of removing the ghost issues as presented above largely depends on the linearized theory. This procedure can nonetheless be performed order by order in the expansion in powers of $h_{\mu\nu}$, removing the ghost pole at each order. Although ghost issues could still reappear non-perturbatively, perturbative theory is usually enough for most practical applications.

As pointed out before, we must also stress that evading the Ostrogradsky instability does not guarantee that other types of ghosts would not originate from some yet unknown mechanism, even when all curvature invariants are present. In fact, if the full theory $S_\infty$ contains more than one pole, one of them is necessarily a ghost should $\mathcal F$ be analytic\footnote{An example of analytic $\mathcal F$ includes the polynomial function $\mathcal F(z) = \sum_{n=0}^N c_n z^n$. On the other hand, an instance of a non-analytic generalized wave operator can be given by $\mathcal F(z) = (1-\alpha\log(z))z$, which appears when one-loop corrections to general relativity are considered \cite{Kuntz:2017pjd}.} \cite{Barnaby:2010kx}. This is easy to see by calculating the residue of the poles in the full theory. Suppose there are $M$ poles (labeled by the index $a$ below) in the propagator of the full theory. Then their residues are given by \cite{Barnaby:2010kx}
\begin{equation}
\eta_a = \frac{1}{2\pi i}\oint_{\mathcal C_a}\mathrm{d}^4q \frac{1}{H(q)} = \frac{1}{H'(m_a^2)},
\end{equation}
where
\begin{equation}
H(z) = \frac{\mathcal F(z)}{\Gamma(z)} = \prod\limits_{a=1}^M (z-m_a^2)
\end{equation}
and $\Gamma(z)$ is everywhere non-zero. The contour $\mathcal C_a$ encloses only the pole at $z=m_a^2$. Thus, as long as $\mathcal F$ is analytic, $\eta_a$ must change sign for $M>1$. In our bottom-up approach, $\mathcal F$ is analytic by construction as it is defined as an infinite series, leading to the conclusion that the scalar degree of freedom is likely to be a byproduct of the truncation as well, otherwise it would be itself a ghost should it exist in the full theory $S_\infty$. While a single massive particle whose decay could produce massive scalars and massless gravitons can very well exist in $S_\infty$, it is clearly impossible to figure out the spectrum of the theory by having at our disposal only an infinite series. If we want to keep the scalar field in the spectrum while insisting that a fundamental theory of quantum gravity can be obtained by the bottom-up construction of the infinite series \eqref{eq:full}, then there is no option other than looking for a non-analytic $\mathcal F$. The most obvious way to make the series \eqref{eq:full} non-analytic is by including negative powers of the curvature in the action, which forces the appearance of poles in $\mathcal F$. It would then be possible to have multiple poles in the propagator with no ghosts in the spectrum, leading to a theory that is both renormalizable and ghost-free, while keeping interesting solutions such as the scalar degree of freedom responsible for Starobinsky inflation.

Alternatively, should we not be interested in Starobinsky inflation, we can stick to the supposedly simpler series \eqref{eq:full} with no negative power of the curvature and project out all undesired modes with the exception of the graviton by means of an integration contour following the same procedure as before. We must note that in the theory \eqref{eq:hd}, it is very natural to pick up a contour that does not enclose the Ostrogradsky ghost because, as we have seen, this ghost is a byproduct of the truncation. On the other hand, the removal of a non-Ostrogradsky ghost in a theory with infinitely many curvature invariants by means of an integration contour might seem less natural, but it is still a legitimate procedure as we always get to choose the boundary conditions. 

In this short paper, we argued that Ostrogradsky ghosts that haunt higher-derivative gravity are actually just artificial byproducts of the truncation of the full theory \eqref{eq:full}, which contains infinitely many curvature invariants. We showed how these fictitious ghosts can be projected out in the truncated theory with the help of an integration contour defined in the Fourier space and which is part of the definition of the theory itself. This allows one to use higher-derivative gravity to study quantum gravity without facing stability or unitarity issues. We also discussed the possibility of building a ghost-free theory containing infinite curvature invariants in its action, without having to get rid of the scalaron responsible for Starobinsky inflation. A more detailed analysis of this last possibility is however needed.

\noindent{\it Acknowledgments:}
This work was supported by the National Council for Scientific and Technological Development -- CNPq (Brazil) under grant number 155342/2018-5 and by the INFN grant FLAG.


\bigskip{}

\end{document}